\documentstyle[12pt]{article}
\normalsize

\let\oldtheequation=\theequation
\def\doteqs#1{\setcounter{equation}{0}
            \def\theequation{{#1}.\oldtheequation}}

\newcounter{sxn}
\def\sx#1{\addtocounter{sxn}{1} \bigskip\medskip \goodbreak \noindent{\large\bf
\centerline{\thesxn.~~#1}} \nobreak \medskip}
\def\sxn#1{\sx{#1} \doteqs{\thesxn}}

\newcounter{axn}
\def\ax#1{\addtocounter{axn}{1} \bigskip\medskip \goodbreak \noindent{\large\bf
{\Alph{axn}.~~#1}} \nobreak \medskip}
\def\axn#1{\ax{#1} \doteqs{\Alph{axn}}}

\def\br{\begin{thebibliography}{abc}}
\def\er{\end{thebibliography}}

\date{}

\begin{document}
\bibliographystyle{unsrt}
\footskip 1.0cm
\thispagestyle{empty}
\begin{flushright}
ISU-NP-94-13\\
DECEMBER 1994\\
\end{flushright}
\vspace*{10mm}
\begin{center} {\LARGE ISGUR-WISE FUNCTION AND $V_{CB}$\\
FROM BETHE-SALPETER EQUATIONS\\}
\vspace*{10mm}
{\large   A. Abd El-Hady, K. S. Gupta, \\
          A. J. Sommerer, J. Spence and J. P. Vary. \\ }
\vspace*{10mm}
{\it Department of Physics and Astronomy,\\
Iowa State University ,\\
Ames, IA 50011, USA}.\\
\newcommand{\bc}{\begin{center}}
\newcommand{\ec}{\end{center}}
\ec
\vspace*{5mm}

\normalsize
\centerline{\bf ABSTRACT}

\vspace*{5mm}

	We calculate the Isgur-Wise function from the solutions of the
Bethe-Salpeter equations. The shape of the Isgur-Wise function thus
calculated is a prediction of the Bethe-Salpeter equations and does
not depend on undetermined parameters. We develop an analytical
approximation to our Isgur-Wise function in the form $\xi ( \omega) =
\eta [ 1 - \frac{{\rho^2}}{\eta} (\omega - 1) + a (\omega - 1)^{3/2}]$
where $\rho^2 = 1.279$, $a = .91$, $\eta = .9942$ and $\omega$ is the
recoil velocity. The Isgur-Wise function is then used to obtain $V_{cb}$
from the recent experimental data of ${\bar B} \to D^* \ell \bar\nu$ decay.
Our best estimate of $V_{cb}$ is $(34.7 \pm 2.5) \times 10^{-3}$, which is
comparable to some of the latest estimates in the literature.
\hfill
\newpage
\newcommand{\be}{\begin{equation}}
\newcommand{\ee}{\end{equation}}

\baselineskip=24pt
\setcounter{page}{1}
\sxn{INTRODUCTION}

	Heavy quark effective theory (HQET) \cite{isg,vol,georgi,neu}
has opened a new window for the
determination of the Cabibbo-Kobayashi-Maskawa (CKM) matrix elements
\cite{neu,neub,stone}. In
particular, determination of $V_{cb}$ using theoretical predictions of HQET and
experimental measurement of the differential decay rate of the exclusive
semileptonic decay ${\bar B} \to D^* \ell \bar\nu$ has received a great deal of
attention. With the recent high-precision data available from several
experimental groups \cite{arg,cle,cleo},
it is now possible to determine the value of $V_{cb}$ with
reasonable accuracy.

	In order to extract $V_{cb}$ from the experimental data, one needs to
calculate hadronic form factors which include nonperturbative effects. HQET has
vastly simplified these calculations and only one universal function,
called the Isgur-Wise (IW) function \cite{isg,neu,rieck},
 has been shown to play a central role in
many calculations involving decay of heavy mesons. The calculation of the IW
function however is model dependent and several different parameterizations
for it have been used in the literature \cite{neu,stone}.
After a particular parameterization is chosen,
one typically fits the experimental data to extract the unknown parameter(s) in
the IW function as well as $V_{cb}$ from the experimental data.

	In this paper we use a covariant reduction of the Bethe-Salpeter
equation (BSE) \cite{bs} to calculate the IW function \cite{kugo}.
The BSE was solved numerically and the
parameters appearing in it (namely the quark masses, string tension and
the running coupling strength for one-gluon exchange) were determined
by fitting the calculated spectrum to the observed masses
of more than 40 mesons \cite{jpv,ala}. The resulting meson mass spectrum
agrees very well with the experimental data.
Once the parameters are thus
fixed, the meson wave functions from the BSE can be used to predict physical
observables. In particular, the IW function may then be evaluated from the
wavefunctions of the BSE and would represent a {\it prediction}
independent of any undetermined parameters. Knowing
the shape of the IW function we can then extract $V_{cb}$ from the experimental
data

	This paper is organized as follows: in Section 2 we recall the main
ingredients for the calculation coming from HQET. Section 3 describes the
salient features of the BSE and describes the meson wavefunctions of interest.
In Section 4 we calculate the IW function and extract $V_{cb}$ from the
experimental data. We conclude the paper in Section 5 with a brief outline of
future work.

\sxn{HEAVY MESON DECAY FORM FACTORS}

	The hadronic matrix elements for the decay
${{\bar B} \to D^*\,\ell\,\bar\nu}$ take a simple form when described in the
context of HQET. Such decays are mediated by heavy quark currents \\
$V_\mu = \bar c {\gamma}_\mu b$ and
$A_\mu =  \bar c {\gamma}_\mu {\gamma}_5 b$ and the corresponding matrix
elements are in general described in terms of four form factors
\cite{isg,rieck}
denoted by $\xi_i$, $i = 1,2,3,4$ :
\begin{eqnarray}
   \langle D^*(v',\epsilon) | V^\mu | {\bar B}(v) \rangle &=&
   i \sqrt{m_B m_{D^*}} \xi_1 (\omega)
\epsilon^{\mu\nu\alpha\beta}\epsilon_\nu^*\,
    v'_\alpha\,v_\beta \,  \nonumber\\
   \langle D^*(v',\epsilon) | A^\mu | {\bar B}(v) \rangle &=&
   \sqrt{m_B m_{D^*}} \xi_2(\omega)\,(\omega +1)\,\epsilon^{\ast\mu} \!-\!
    (\xi_3(\omega)\,v^\mu + \xi_4(\omega)v'^\mu)
    \epsilon^* .v,
\end{eqnarray}
where $\omega = v.v^{\prime}$, $v$ and $v^{\prime}$ being the velocities of
${\bar B}$ and $D^*$ meson respectively.

	In the limit where masses of the heavy quarks tend to infinity, the
form factors $\xi_i$ satisfy the conditions
\begin{equation}
\xi_1 = \xi_2 = \xi_4 \equiv \xi (v.v'), ~~~~ \xi_3 = 0,
\end{equation}
where $\xi(v.v')$ is a single universal function, called the IW function
\cite{isg,neu}.
In the limit of infinitely heavy quark masses
the IW function is normalized to unity at zero recoil, i.e.
$\xi(1) = 1$.

	The IW function can be related to the overlap integral of
normalized meson wave functions in the infinite momentum frame. If $\psi_{l
\bar B}$ and $\psi_{l D^*}$ denote the wavefunctions of the light degrees of
freedom in $\bar B$ and $D^*$ mesons respectively, then the IW function
can be written as
\begin{equation}
\xi (\omega) = \big ({\frac{2}{\omega + 1}}\big )^{1/2}
\int  \psi^*_{l D^*}(v^{\prime}) \psi_{l \bar B}(v) d^3x.
\end{equation}

	In the heavy quark limit close to $\omega = 1$, the IW function has the
form
\begin{equation}
\xi (\omega) = 1 - \rho^2 (\omega - 1) + O[(\omega - 1)^2],
\end{equation}
where $\rho^2$ is the slope of the Isgur-Wise function at $\omega = 1$.

	The differential decay rate for the process discussed above is given
by \cite{neub}
\begin{eqnarray}
   {{\rm d}\Gamma({\bar B}\to D^*\ell \bar\nu)\over{\rm d}\omega}
   &=& {G_F^2\over 48\pi^3}\,(m_{\bar B}-m_{D^*})^2\,m_{D^*}^3\,\eta_A^2\,
    \sqrt{\omega^2-1}\,(\omega+1)^2 \nonumber\\
   \phantom{ \bigg[ }
   &&\times \bigg[ 1 + {{4 \omega} \over {\omega+1}}
\,{1-2\omega r+r^2\over(1-r)^2}
    \bigg]\,|\,V_{cb}|^2\,{\xi}^{\,2}(\omega),
\end{eqnarray}
where $r = {{m_{D^*}} \over {m_{\bar B}}}$ and $\eta_A$ is a constant which
is present due to a finite renormalization of the axial vector current.

	It is clear from the above expression that the knowledge of the
differential decay rate and of the Isgur-Wise function would allow us to
calculate $V_{cb}$. However, as is evident from Eqn.(2.3), we need to know the
meson wave functions to calculate the Isgur-Wise function. In the next section
we show how to do this using a covariant reduction of the BSE.

\sxn{MESON WAVEFUNCTIONS FROM THE BSE}

	In a previous set of works, reductions of the  BSE has been solved for
$q {\bar q}$ systems \cite{jpv,ala}.
In this section we shall briefly describe our treatment and the wavefunctions
resulting from our calculation.
The actual equations solved to obtain these wavefunctions are given in the
appendix and the reader is referred to the previous works cited above for a
full
treatment.

	The BSE is a covariant four-dimensional wave equation for relativistic
bound states and is very challenging to solve exactly for realistic kernels.
One typically uses several approximations to
reduce it to a solvable form.
We have used a quasipotential equation framework to
reduce the BSE to the three dimensional integral equation introduced in
\cite{jpv} and have retained only the ladder diagram component of the full BSE
kernel. Under this approximation the interaction kernel contains a
one-gluon exchange term $V_{oge}$, to which we add a phenomenological,
long-range confining potential,
$V_{con}$. The interaction kernel that we use is thus derived from the
potential
\begin{equation}
V_{oge}+V_{con}={4\over 3}\alpha_s{\gamma_\mu\otimes\gamma_\mu\over
{(q-q')^2}}
+\sigma{\rm\lim_{\mu\to 0}}{\partial^2\over\partial\mu^2}
{{\bf 1}\otimes{\bf 1}\over-(q-q')^2+\mu^2}.
\end{equation}
Here, $\alpha_s$ is the strong coupling, which is weighted by the meson color
factor of ${4\over 3}$, and the string tension $\sigma$ is the strength of
the confining part of the interaction.

	We also take into account the running of the strong coupling constant.
Specifically, we use the form
\begin{equation}
\alpha_s(Q^2)={4\pi\alpha_s(\mu^2)\over 4\pi+\beta\alpha_s(\mu^2){\rm
ln}\bigl({Q^2/\mu^2}\bigr)},
\end{equation}
where
\begin{equation}
Q^2 = \gamma^2 M_{meson}^2 + \beta^2,
\end{equation}
and where $\gamma$ and $\beta$ are parameters determined by a fit to the meson
spectrum. In the above equation $\mu$ is taken to be the mass
of $Z$ boson and $\alpha_s (\mu^2)$
is correspondingly chosen to be equal to 0.12 based
on experimental measurements.

	In our formulation of BSE there are seven parameters :
four masses,
$m_{u}$=$m_{d}$, $m_c$, $m_{s}$, $m_{b}$; the string tension $\sigma$, and the
parameters $\gamma$ and $\beta$ used to govern the running of the  coupling
constant. These parameters were determined by fitting
the meson masses calculated from the BSE to the observed spectrum.
We utilize this BSE model for the mesons to evaluate the meson wavefunctions.
The wavefunctions for $B$ and $D^*$
mesons are shown in Figs. (1) and (2) respectively. For the purpose of
further calculations we have also obtained analytic representations of these
wavefunctions :
\begin{equation}
\psi (x,y,z) = ({{(2^{1/n} \lambda )^3 n} \over {\Gamma({{3} \over
{n}}})})^{\frac{1}{2}} e^{- {\lambda}^n (x^2 + y^2 + z^2)^{\frac{n}{2}}},
\end{equation}
where, $\lambda = .59$ GeV, $n = 1.16$ for the $\bar B$ meson and
$\lambda = .42$ GeV, $n = 1.52$ for the $D^*$ meson.
We have plotted the numerical wavefunctions and their analytic representations
together to display the accuracy of the latter. For convenience,
we shall use these analytical expressions for subsequent calculations in this
paper.

\sxn{ISGUR-WISE FUNCTION AND $V_{cb}$}

	In this section we shall describe our results using the ingredients
that have been presented above. Using Eqn.(2.3) and taking into account the
relativistic boost (assumed along the $z$ direction),
the IW function can be written as \cite{rieck}
\begin{equation}
\xi (\omega) = \big ({\frac{2}{\omega + 1}}\big )^{1/2}
\int  \psi^*_{l D^*}(x,y,\omega z) \psi_{l \bar B}(x,y,z)
e^{i E z\sqrt{w^2 - 1}} d^3x,
\end{equation}
where $E$ is the mass of the light component of the
$D^*$ meson in the rest frame of
the $B$ meson.

 We use the wavefunctions derived from the BSE to evaluate the IW
function and we emphasize again that our IW function involves no
additional parameter fitting. The plot of our IW function is shown in
Fig. (3). It is interesting to note that since the wave functions
used are coming from  the BSE
which is solved independent of the heavy quark approximation, the
IW function does not go to unity
at zero recoil. The deviation of the IW function from unity  at zero
recoil can be attributed to the finite mass corrections which are
incorporated in our BSE model. We have also
obtained an analytical representation of the IW function that is
plotted in Fig.(3). This is given by
\begin{equation}
\xi ( \omega) = \eta [ 1 - \frac{{\rho^2}}{\eta} (\omega - 1) + a (\omega -
1)^{3/2}],
\end{equation}
where $\rho^2 = 1.279$, $a = .91$ and $\eta = .9942$.

	We can now use the IW function as calculated above to extract
$V_{cb}$
from the experimental data for ${{\bar B} \to D^*\,\ell\,\bar\nu}$
decay. We have used the data from ARGUS 93 \cite{arg},
CLEO 93 \cite{cle} and CLEO 94 \cite{cleo} and the corresponding fits
to these data using our IW function are shown in the Figs. (4), (5) and (6)
respectively.

In Table 1 we present the different values of $V_{cb}$
(in the units of $10^{-3}$)
obtained from these experimental data (for each set of experimental data
we have
shown two values of $V_{cb}$, one obtained from the numerical calculation
of the
IW function and the other from its analytic representation as given in Eqn.
(4.2).)
\begin{table}\centering
\caption{Values of $V_{cb}$ in the units of $10^{-3}$
from different sets of experimental data. The first row
indicates the values of $V_{cb}$ obtained by using the IW function
that was calculated numerically. The second row indicates the corresponding
values of $V_{cb}$ obtained by using an analytical approximation to the
IW function as given in Eqn. (4.2).}
\bigskip
\begin{tabular}{|l|c||c||c|}
\hline
   & ARGUS 93& CLEO 93& CLEO 94\\
\hline\hline
${\rm From ~Numerical~  IW~Function}$  &$(35.8\pm 8.3)$
&$(34.8\pm 6.1) $&$(34.7\pm 2.5) $\\
\hline\hline
${\rm From~ Analytical~ IW ~ Function}$  &$(35.9\pm 8.3) $
&$(34.8\pm 6.1) $&$(34.7\pm 2.5) $\\
\hline\hline
\end{tabular}\\
\end{table}
The uncertainties in the values of $V_{cb}$ tabulated above are
determined by fitting to the two extreme values appearing on the experimental
error bars. The different values thus obtained are consistent with each other
and are comparable to some of the latest estimates of $V_{cb}$
\cite{neu,neub,stone}.

\sxn{CONCLUSION}

	We have used the solutions of the BSE to predict the shape of the IW
function. Such a prediction is independent of any undetermined parameter.
The solution of BSE does not depend on the heavy quark approximation and the
corresponding IW function calculated from its solutions incorporates some of
the
finite mass corrections. This is evident from the fact that in our calculation
the IW function deviates from unity at zero recoil. The analytical
representation
of the IW function shows an interesting power law behaviour which is different
from the usual representations available in the literature. Using the IW
function we have calculated $V_{cb}$ from the latest set of experimental data.
Our best estimate of $V_{cb} = (34.7 \pm 2.5) \times 10^{-3}$ is comparable to
the latest estimates of $V_{cb}$ available in the literature
\cite{neu,neub,stone}.

	From the knowledge of the solutions of $V_{cb}$ it should be possible to
calculate the four form factors in Eqn. (2.1) directly without using the heavy
quark limit. Such a calculation would provide
an interesting test of the accuracy and
consistency of the heavy quark approximation and is currently being pursued.

\bigskip

\begin{center}
{\bf ACKNOWLEDGEMENTS}
\end{center}

We thank A.P.Balachandran, C.Benesh,
M.Burkardt, M.Harada, G.Moneti, A.Petridis, J.W.Qiu, J.Schechter,
S.Stone and B.-L.Young for discussions. K.S.G. acknowledges the
hospitality of the Physics
Department of Syracuse University where a part of this work was completed.
This work was supported in part by the US Department of Energy, Grant
No. DE-FG02-87ER40371, Division of High Energy and Nuclear Physics.

\begin{center}
{\bf APPENDIX}
\end{center}

\axn{BETHE SALPETER EQUATIONS}

	In this appendix we present the BSE that have been solved. This is done
for the sake of completeness and the reader should refer to \cite{jpv,ala} for
further details.

	The BSE for the $0^-$ channel is given by
\begin{equation}
[(E_1 + E_2)^2 - E^2]\psi^0(q) = {{E} \over {\pi q}}[I_0^{con} \psi^0(q^{'}) +
                                   I_0^{oge} \psi^0(q^{'})],
\end{equation}
and those for the $1^-$ channel are given by
\begin{equation}
[(E_1 + E_2)^2 - E^2]\psi^{12}(q) ={E \over {\pi q}}[(I_1^{con} +I_1^{oge})
                      \psi^{12}(q^{'}) + (I_2^{con} + I_2^{oge})
\psi^{34}(q^{'})],
\end{equation}
and
\begin{equation}
[(E_1 + E_2)^2 - E^2]\psi^{34}(p) ={E \over {\pi q}}[(I_3^{con} +I_3^{oge})
                      \psi^{12}(q^{'}) + (I_4^{con} + I_4^{oge})
\psi^{34}(q^{'})].
\end{equation}
The symbols appearing in the above equations are defined as :
$$
I_0^{con} = \sigma{\rm\lim_{\mu\to 0}}{\partial^2\over\partial\mu^2}
            \int_0^{\infty} d
q^{'}q^{'}[Q_0(z^{'})T_0^{con}+Q_0^1(z^{'})T_1^{con}],
$$
$$
I_1^{con} = \sigma{\rm\lim_{\mu\to 0}}{\partial^2\over\partial\mu^2}
            \int_0^{\infty} d
q^{'}q^{'}[Q_0(z^{'})T_1^{con}+Q_0^1(z^{'})T_2^{con}],
$$
$$
I_2^{con} = \sigma{\rm\lim_{\mu\to 0}}{\partial^2\over\partial\mu^2}
            \int_0^{\infty} d q^{'}q^{'}[Q_1^3(z^{'})T_2^{con}] ,
$$
$$
I_3^{con} = \sigma{\rm\lim_{\mu\to 0}}{\partial^2\over\partial\mu^2}
            \int_0^{\infty} d q^{'}q^{'}[Q_1^3(z^{'})T_2^{con}] ,
$$
$$
I_4^{con} = \sigma{\rm\lim_{\mu\to 0}}{\partial^2\over\partial\mu^2}
            \int_0^{\infty} d
q^{'}q^{'}[Q_1(z^{'})T_1^{con}+Q_1^2(z^{'})T_0^{con}],
$$
$$
I_0^{oge} = {{q \alpha_s} \over {3}}
            \int_0^{\infty} d q^{'}q^{'}[Q_0(z)T_0^{oge}+Q_0^1(z)T_1^{oge}],
$$
$$
I_1^{oge} = {{q \alpha_s} \over {3}}
            \int_0^{\infty} d q^{'}q^{'}[Q_1(z)T_3^{oge}+Q_1^1(z)T_4^{oge}],
$$
$$
I_2^{oge} = {{q \alpha_s} \over {3}}
            \int_0^{\infty} d q^{'}q^{'}[Q_1^3(z)T_5^{oge}],
$$
$$
I_3^{oge} = {{q \alpha_s} \over {3}}
            \int_0^{\infty} d q^{'}q^{'}[Q_1^3(z)T_7^{oge}],
$$
$$
I_4^{oge} = {{q \alpha_s} \over {3}}
            \int_0^{\infty} d q^{'}q^{'}[Q_1(z)T_2^{oge}+Q_1^2(z)T_1^{oge}],
$$
$$
T_0^{con} = {-1 \over {2 (E_1 E_2 E_1^{'} E_2^{'})}}
            (E_1 E_2 E_1^{'} E_2^{'} + q^2 {q^{'}}^2),
$$
$$
T_1^{con} = {1 \over {2 (E_1 E_2 E_1^{'} E_2^{'})}}
            q q^{'} (E_1 E_1^{'} + E_2 E_2^{'}),
$$
$$
T_2^{con} = {1 \over {2 (E_1 E_2 E_1^{'} E_2^{'})}}
            (E_1 E_2 E_1^{'} E_2^{'} - q^2 {q^{'}}^2),
$$
$$
T_0^{oge} = {1 \over {4 (E_1 E_2 E_1^{'} E_2^{'})}}
            (E_1 E_2 E_1^{'} E_2^{'} + q^2 {q^{'}}^2 + 3q^2 E_1^{'} E_2^{'}
            + 3{q^{'}}^2 E_1 E_2),
$$
$$
T_1^{oge} = {1 \over {4 (E_1 E_2 E_1^{'} E_2^{'})}}
            (E_1 E_2 E_1^{'} E_2^{'} + q^2 {q^{'}}^2 + q^2 E_1^{'} E_2^{'}
            +{q^{'}}^2 E_1 E_2),
$$
$$
T_2^{oge} = {1 \over {4 (E_1 E_2 E_1^{'} E_2^{'})}}
            q q^{'} (E_1 E_1^{'} + E_2 E_2^{'} + E_1^{'} E_2 + E_1 E_2^{'}),
$$
$$
T_3^{oge} = {1 \over {4 (E_1 E_2 E_1^{'} E_2^{'})}}
            4q q^{'} (E_1 E_1^{'} + E_2 E_2^{'}) ,
$$
$$
T_4^{oge} = {1 \over {4 (E_1 E_2 E_1^{'} E_2^{'})}}
            (E_1 E_2 E_1^{'} E_2^{'} + q^2 {q^{'}}^2 - q^2 E_1^{'} E_2^{'}
            -{q^{'}}^2 E_1 E_2),
$$
$$
T_5^{oge} = {-1 \over {4 (E_1 E_2 E_1^{'} E_2^{'})}}
            (E_1 E_2 E_1^{'} E_2^{'} - q^2 {q^{'}}^2 + q^2 E_1^{'} E_2^{'}
            -{q^{'}}^2 E_1 E_2),
$$
$$
T_6^{oge} = {1 \over {4 (E_1 E_2 E_1^{'} E_2^{'})}}
            q q^{'} (E_1- E_2) (E_1^{'}- E_2^{'}) ,
$$
\begin{equation}
T_7^{oge} = {-1 \over {4 (E_1 E_2 E_1^{'} E_2^{'})}}
            (E_1 E_2 E_1^{'} E_2^{'} - q^2 {q^{'}}^2 - q^2 E_1^{'} E_2^{'}
            +{q^{'}}^2 E_1 E_2),
\end{equation}
where
$$
E_1 = (m_1^2 + q^2)^{\frac{1}{2}},~~~~~E_2 = (m_2^2 + q^2)^{\frac{1}{2}},
$$
$$
E_1^{'}=(m_1^2+{q^{'}}^2)^{\frac{1}{2}},~~~~~E_2^{'}=(m_2^2+{q^{'}}^2)^{
\frac{1}{2}},
$$
$$
z={{q^2 + {q^{'}}^2} \over {2 q q^{'}}},~~~~~z^{'}={{q^2+{q^{'}}^2 +\mu^2}\over
{2qq^{'}}}.
$$
In the above equations, $Q_j$ are the Legendre polynomials of the second kind
and
$$
Q_j^1(z) = z Q_j (z) - \delta_{j 0},
$$
$$
Q_j^2(z) = {1 \over {j + 1}} (jz Q_j (z) + Q_{j - 1}(z)),
$$
$$
Q_j^3(z) = ({j \over {j + 1}})^{\frac{1}{2}} (z Q_j (z) + Q_{j - 1}(z)).
$$

Finally, we also have
$$
\psi_{l=j-1}^j (q) = { {i (-j + 1)} \over { (2 (2j +1))^{\frac{1}{2}}}}
                     (j^{\frac{1}{2}} \psi^{12~j}(q) +
                     (j +1)^{\frac{1}{2}} \psi^{34~j}(q)),
$$
\begin{equation}
\psi_{l=j+1}^j (q) = { {i (-j - 1)} \over { (2 (2j +1))^{\frac{1}{2}}}}
                     (-(j +1)^{\frac{1}{2}} \psi^{12~j}(q) +
                     j^{\frac{1}{2}} \psi^{34~j}(q)).
\end{equation}

$\psi^{12~j}(q)$ and $\psi^{34~j}(q)$ are obtained by numerically solving Eqns.
(A.2) and (A.3). $\psi_{l=j-1}^j (q)$ and $\psi_{l=j+1}^j (q)$ are then
obtained numerically from Eqn. (A.5). These are the wavefunctions for the $D^*$
meson. The $l=2$ component of the $D^*$ meson wavefunction is negligible
in comparison to the $l=0$ component and is neglected. Similarly,
$\psi^0 (q)$ (representing the wavefunction of the ${\bar B}$ meson) is
obtained
by numerically solving Eqn. (A.1). Analytical approximations of $\psi^0 (q)$
and
$\psi_1^1 (q)$ are given in Eqn. (3.4). They are used in Eqn. (4.1) to
calculate
the IW function. Fig. 1 and Fig. 2 show the numerical wavefunctions and their
analytical representations for ${\bar B}$ and $D^*$ mesons respectively.

\newpage

\begin{center}
{\bf FIGURE CAPTIONS}
\end{center}

\noindent
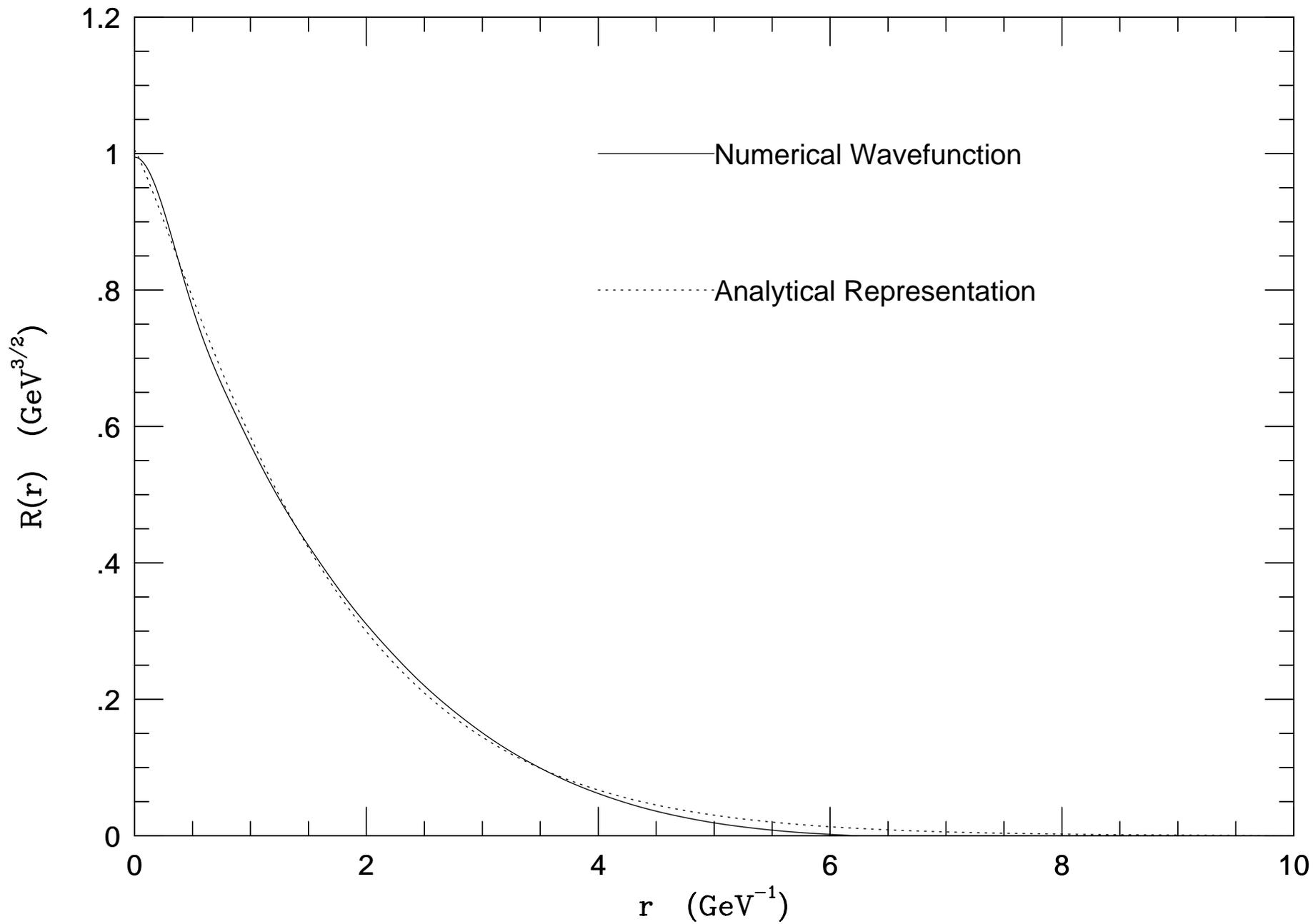
{\bf Figure 1}: Radial wavefunction for the ${\bar B}$ meson.
The solid line represents the numerical solution of the BSE.
This is obtained
from the Fourier transform of $\psi^0 (q)$ appearing in Eqn. (A.1).
The entire Fourier transform of $\psi^0 (q)$ in position space is
normalized
to unity. An analytical approximation to $\psi^0 (q)$ is given in Eqn.
(3.4)
(with $\lambda=0.59$ GeV and $n=1.16$) and is represented by the dashed
line.

\noindent
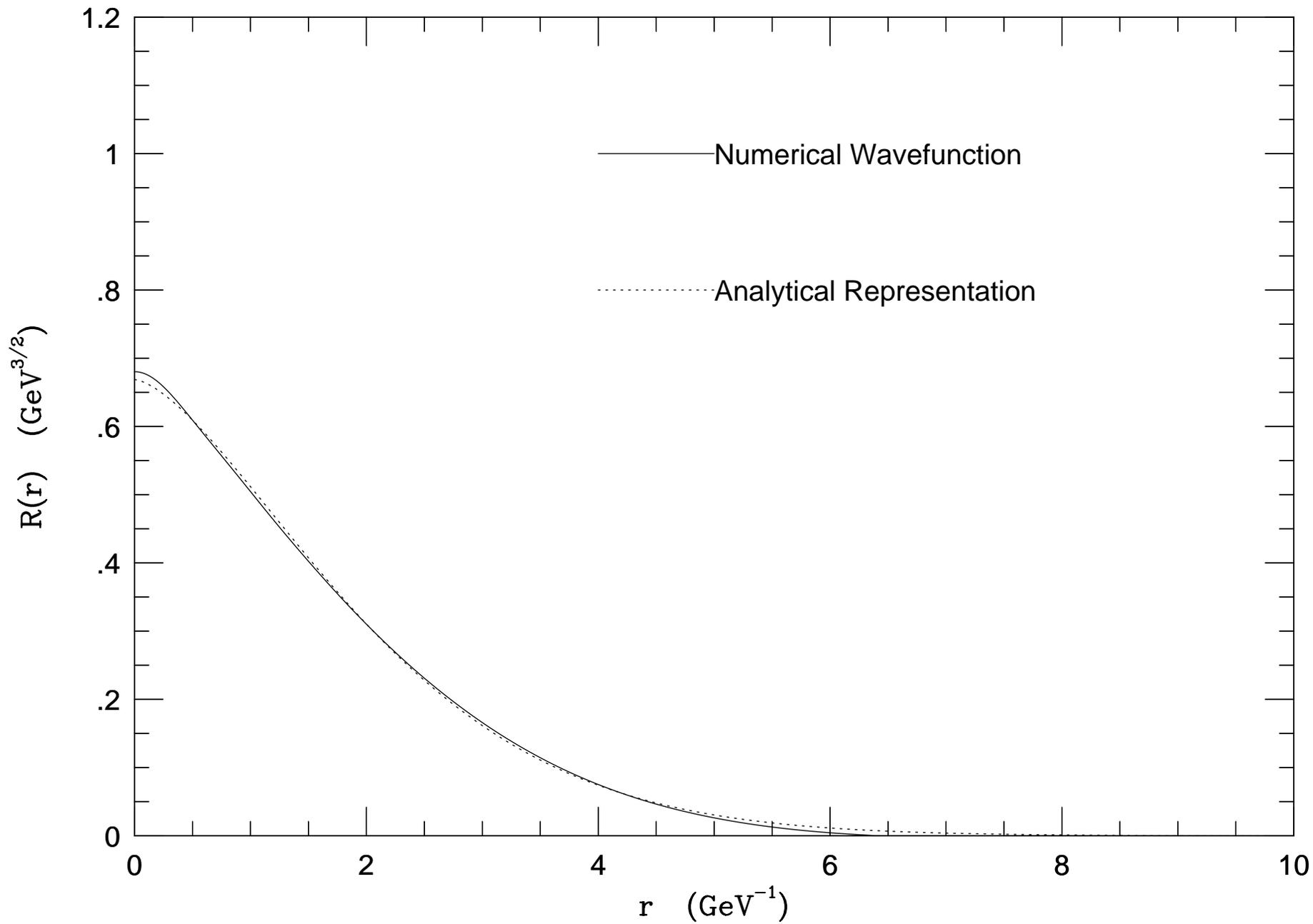
{\bf Figure 2}: Radial wavefunction for the ${D^*}$ meson.
The solid line represents the numerical solution of the $l=0$ component.
This is obtained from the Fourier transform of $\psi^1_0 (q)$ appearing
in Eqn. (A.5).
The entire Fourier transform of $\psi^1_0 (q)$ in position space is
normalized to unity. An analytical approximation to $\psi^1_0 (q)$ is
given in Eqn. (3.4)
(with $\lambda=0.42$ GeV and $n=1.52$) and is represented by the
dashed line.

\noindent
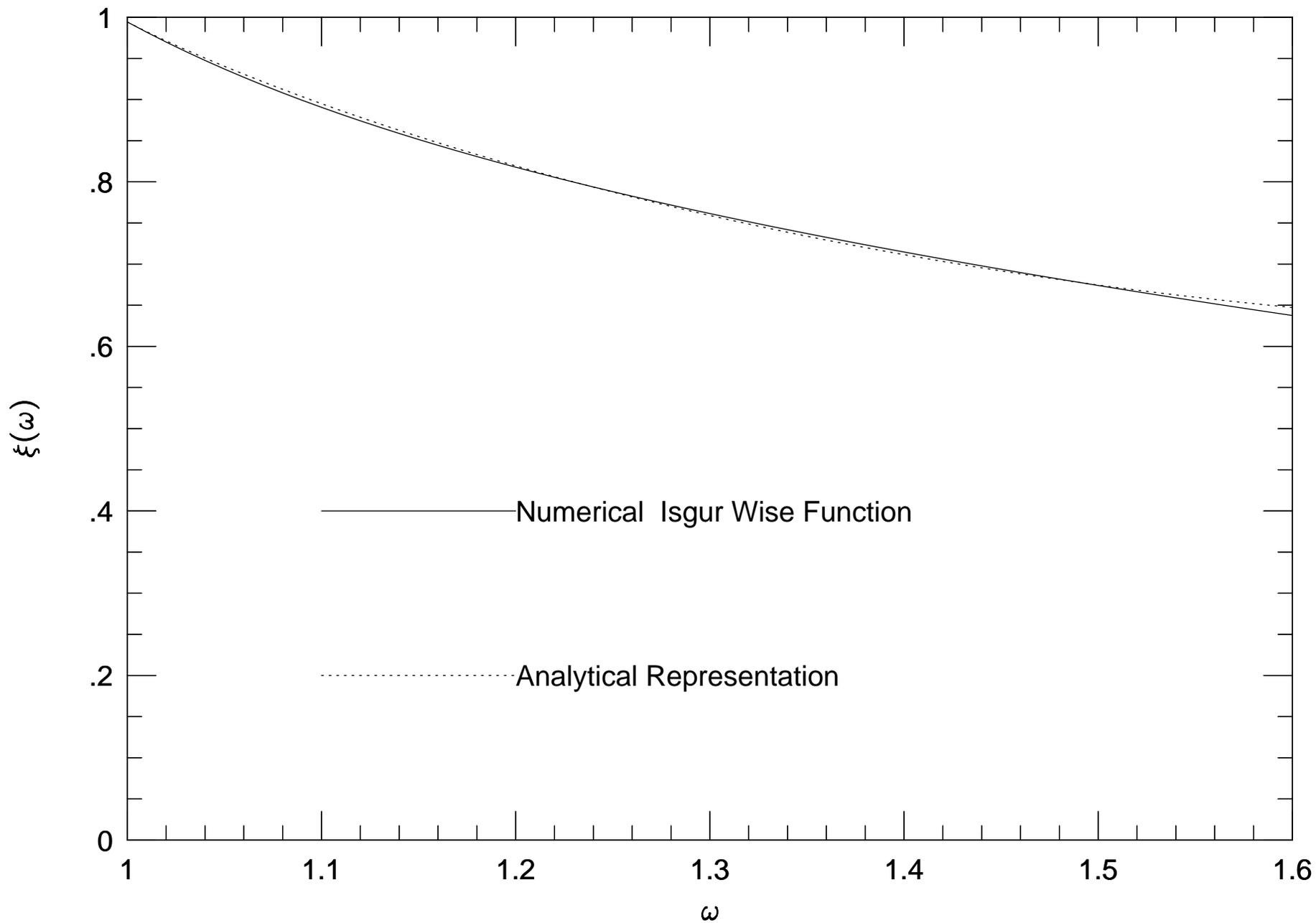
{\bf Figure 3}: The Isgur-Wise function.
The solid line represents the IW function calculated from Eqn.
(4.1). An analytical approximation to the IW function is given in Eqn. (4.2)
and
is represented by the dashed line. The deviation of the IW function from unity
at zero recoil is due to the finite heavy quark mass effects incorporated
in our BSE model.

\noindent
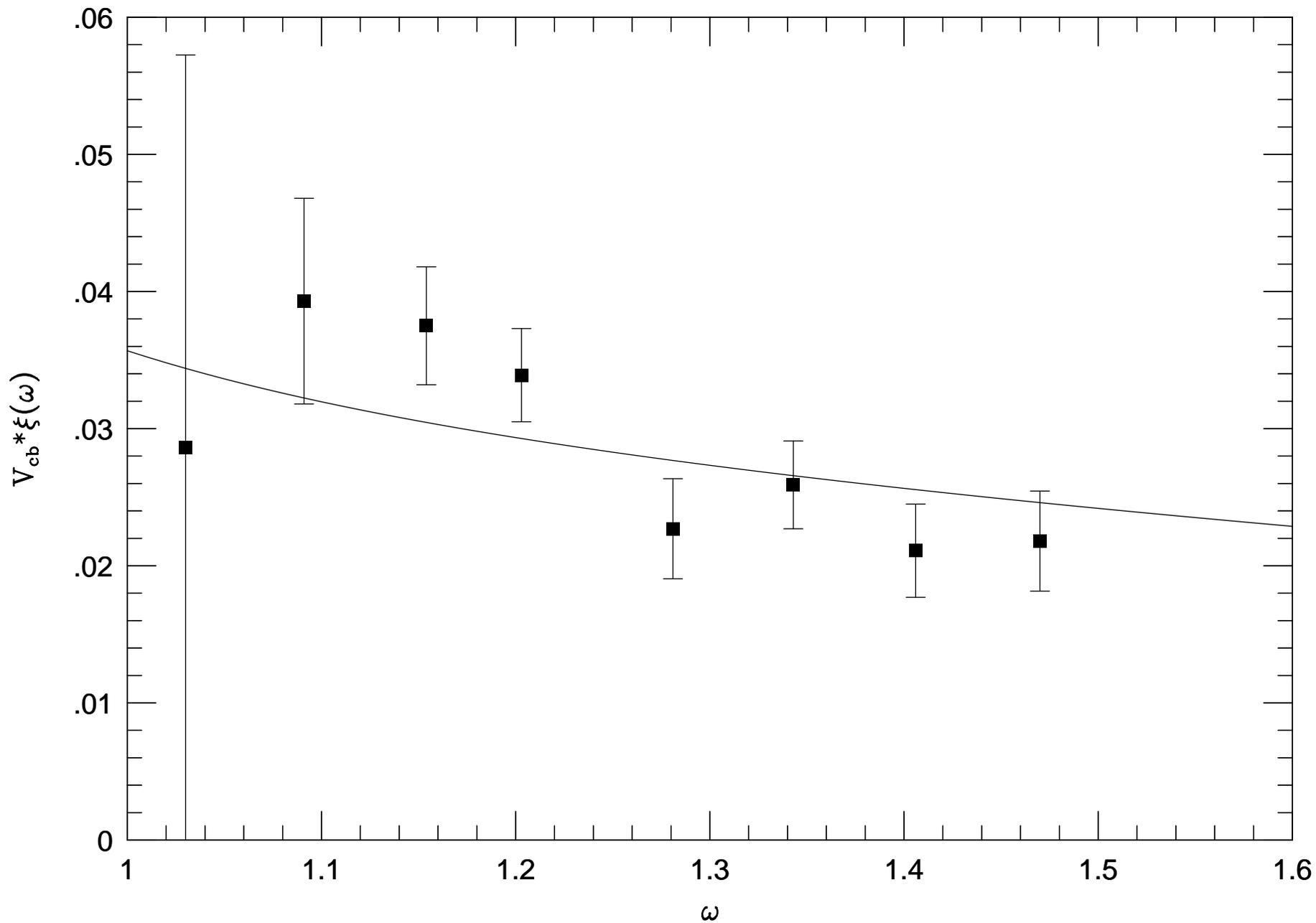
{\bf Figure 4}: Plot of $V_{cb}~ \xi (\omega)$ vs. $\omega$ from ARGUS 93
data of ${\bar B} \to D^* \ell \bar\nu$ decay.
\cite{arg}. The data points for $V_{cb}~ \xi (\omega)$ are extracted
from the experimental result by using Eqn. (2.5). The solid line represents
our best fit to the data.

\noindent
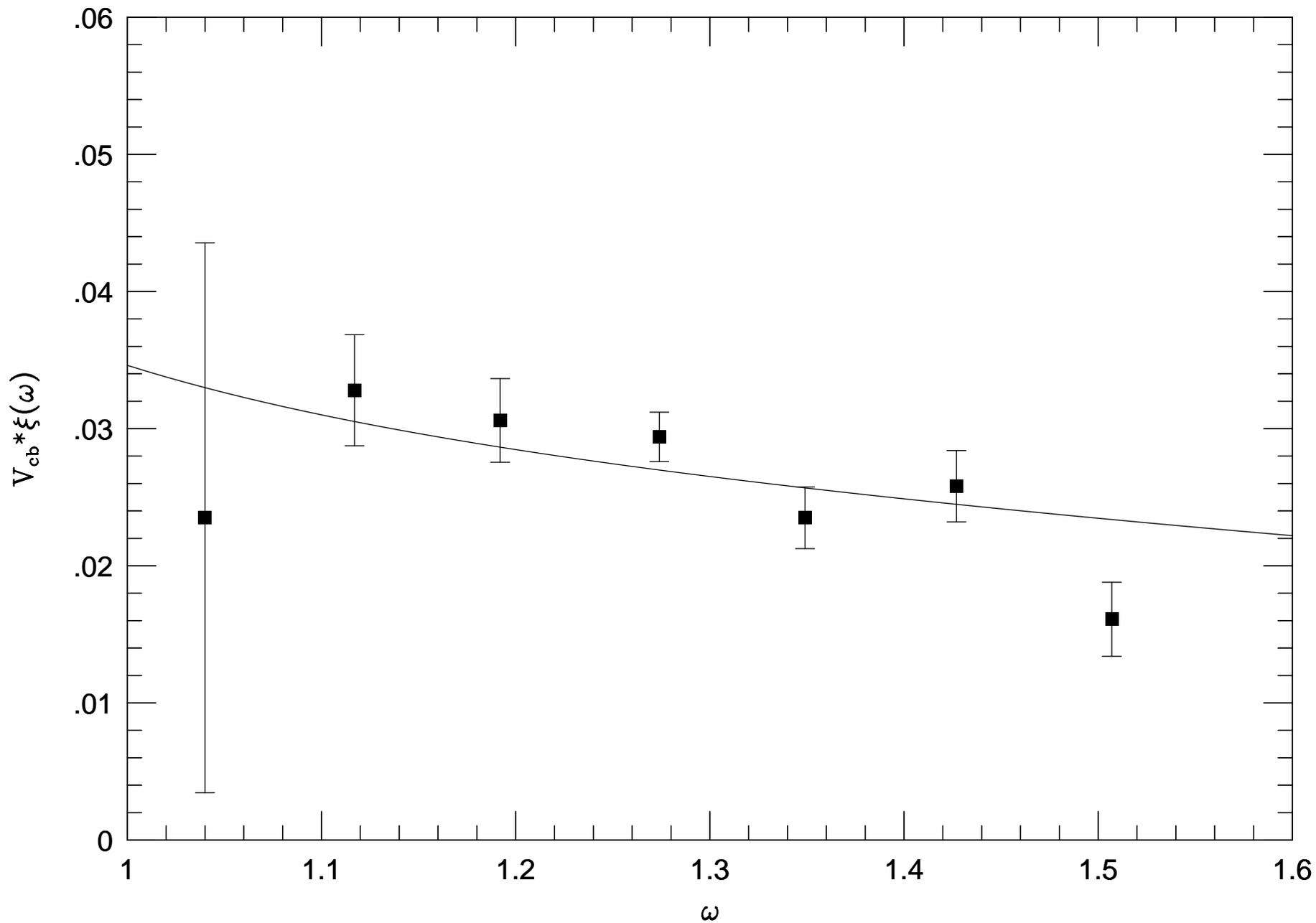
{\bf Figure 5}: Plot of $V_{cb}~ \xi (\omega)$ vs. $\omega$ from CLEO 93
data of ${\bar B} \to D^* \ell \bar\nu$ decay.
\cite{arg}. The data points for $V_{cb}~ \xi (\omega)$ are extracted
from the experimental result by using Eqn. (2.5). The solid line represents
our best fit to the data.

\noindent
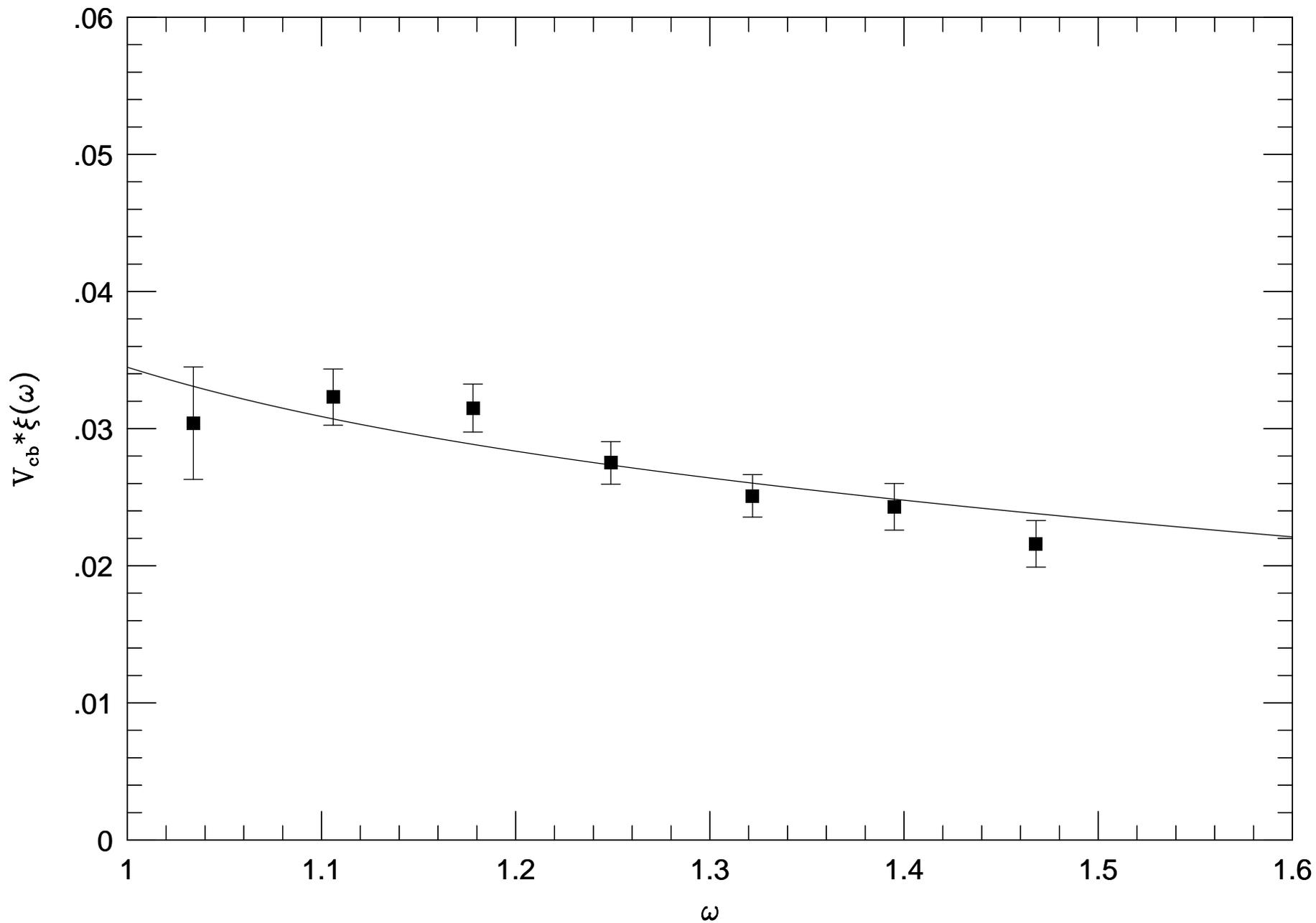
{\bf Figure 6}: Plot of $V_{cb}~ \xi (\omega)$ vs. $\omega$ from CLEO 94
data of ${\bar B} \to D^* \ell \bar\nu$ decay.
\cite{arg}. The data points for $V_{cb}~ \xi (\omega)$ are extracted
from the experimental result by using Eqn. (2.5). The solid line represents
our best fit to the data.


\vfill

\newpage
\begin{thebibliography}{99}

\bibitem{isg}
N. Isgur and M. B. Wise
Phys. Lett. {\bf B232}, 113 (1989); {\bf B237}, 527 (1990).

\bibitem{vol}M. B. Voloshin and M. A. Shifman, Yad. Fiz.
{\bf 47}, 801 (1988) [Sov. J. Nucl. Phys. {\bf 47}, 511 (1988).

\bibitem{georgi}
H. Georgi, Phys. Lett. {\bf B240}, 447 (1990).

\bibitem{neu}
M. Neubert, Phys. Rep. {\bf 245}, 259 (1994) and references therein.

\bibitem{neub}
M. Neubert, Phys. Lett. {\bf B338}, 84 (1994).

\bibitem{stone}
S. Stone, Syracuse University preprint HEPSY 94-5, September 1994.

\bibitem{arg}
A. Albrecht et al., (ARGUS Collaboration), Z. Phys. {\bf C 57},
533 (1993).

\bibitem{cle}
G. Crawford et al., (CLEO Collaboration), to appear in the Proceedings of 16th
International Symposium of Lepton and Photon Interactions, Ithaca, N.Y. 1993.

\bibitem{cleo}
B. Barish et al., (CLEO Collaboration), Cornell Nuclear Studies Wilson Lab
preprint, HEPEX 9406005 (1994).

\bibitem{rieck}
M. Neubert and V. Rieckert, Nucl. Phys. {\bf B382}, 97 (1992).

\bibitem{bs}
E. E. Salpeter and H. A. Bethe, Phys. Rev. {\bf 84}, 1232 (1951).

\bibitem{kugo}
T. Kugo, M. G. Mitchard and Y. Yoshida, Prog. Theo. Phys. {\bf 91},
521 (1994).

\bibitem{jpv}
A. J. Sommerer, J. R. Spence and J. P. Vary, Phys. Rev. {\bf C 49},
513 (1994).

\bibitem{ala}
A. J. Sommerer, A. Abd El-Hady, J. Spence and J. P. Vary,
Iowa State University preprint ISU-NP-94-07, October 1994.

\end{thebibliography}
\end{document}